\documentclass[10pt]{article}

\usepackage{amsmath}
\usepackage{amssymb}

\usepackage{graphicx}

\usepackage{cite}

\usepackage{color} 

\usepackage[labelfont=bf,labelsep=period,justification=raggedright]{caption}

\makeatletter
\renewcommand{\@biblabel}[1]{\quad#1.}
\makeatother

\date{}

\begin{document}

\begin{flushleft}
{\Large
\textbf{Cliques in complex networks reveal link formation and community evolution}
}
\\
Zhen Liu$^{1,2,\ast}$, 
Jia-Lin He$^{1}$, 
Jaideep Srivastava$^{2}$
\\
\bf{1} Web sciences center, School of computer science and engineering, University of electronic science and technology of China, ChengDu, SiChuan, China
\\
\bf{2} Department of computer science and engineering, University of Minnesota, Minneapolis, Minnesota, USA
\\

$\ast$ E-mail: quake@uestc.edu.cn
\end{flushleft}

\section*{Abstract}
Missing link prediction in indirected and un-weighted network is an open and challenge problem which has been studied intensively in recent years. In this paper, we studied the relationships between community structure and link formation and proposed a Fast Block probabilistic Model(FBM). In accordance with the experiments on four real world networks, we have yielded very good accuracy of missing link prediction and  huge improvement in computing efficiency compared to conventional methods. By analyzing the mechanism of link formation, we also discovered that clique structure plays a significant role to help us understand how links grow in communities. Therefore, we summarized three principles which are proved to be able to well explain the mechanism of link formation and network evolution from the theory of graph topology. 
\section*{Author Summary}

\section*{Introduction}
Recently, the study of link prediction has attracted much attention from disparate scientific communities. In the theoretical aspect, accurate prediction indeed gives evidence to some underlying mechanisms that drive the network evolution\cite{1}. Moveover, it is very possible to build a fair evaluation platform for network modeling under the framework of link prediction, which might be interested by network scientists\cite{2,3}. In the practical aspect, for biological networks such as protein-protein interaction networks and metabolic networks \cite{4,5,6}, the experiments of uncovering new links or interactions are costly, and thus to predict in advance and focus on the links most likely to exist can sharply reduce experimental costs\cite{7}. In the Facebook social network, the link prediction based on supervised random walks has been applied to predict and recommend possible future friends to the current user\cite{8}. link prediction approach is also verified very useful for human preferences recommendation in the field of so called collaborative filtering\cite{9}. In Massively Multiplayer Online Role Playing Game(MMORPG) networks, the studies of link prediction have given an insight to uncover underlying relationships between game players\cite{10,11,12}.
\\
In recent years, there are various research fields developed by the studies of link prediction in complex networks. Missing link prediction, as a fundamental issue, aims at estimating the likelihood of the existence of a link between two nodes in a given network based on the observed links\cite{2,13,14}. Similar to the missing link prediction, spurious link prediction is also carefully studied by some researchers\cite{15}. In online social networks, determining positive and negative links is another interesting problem of link prediction which has aroused people's attention recently\cite{16} because researchers believe that relations such as friendship should be opposite to other relations such as antagonism. On the other hand, since network is very dynamic, given link data for times 1 through T, can we predict the links at time T+1? Therefore, researchers also try to investigate the issue of temporal link prediction\cite{17}. Likewise, the link prediction in multiple networks become possible as well, which means the task to predict links in a network by only using features from other networks\cite{10}.
\\
As for the studying fields mentioned above, various models and methods have been proposed. Some conventional algorithms mainly take the link prediction as a classification issue and therefore apply classical machine learning methods based on node attributes information\cite{18,19,20,21}, and the attributes, for example, would be a people's age, sex, friend prefence, and so on in a social network or a computer's IP adress, MAC address, operating system type, etc in a computer network. However, scientists have found that network's structure information is more powerful than nodes attributes in most contexts and proposed many simple but effective methods merely utilizing local or global structure information of the network, such as the methods of common neighbor, Jaccard, Katz\cite{22}, Adamic Adar\cite{23}, and resource allocation\cite{24}, etc. Recently, there are several hybrid methods that attemt to combine node attributes with structure information appeared in some literatures as well\cite{8,10}, but most of them are applied in some specified domains which mean that domain knowledge must be required.
\\
Since link prediction based on structure features of the network is more general, in this paper, we focus on the issue of missing link prediction in indirected and un-weighted networks by merely using network's topology information. As we believe the community would be the cradle to promate link formation, we present a novel model based on the community structure and statisitc theory. To assess the performance of our model, we evaluate it on four real world networks. Results show that our model improves both the prediction accuracy and computing efficiency. Due to the vital theoretical interest to understand the mechanism of link formation, we evolve a encouraging theory framework to explain this issue which is proved sound by our experiments.

\section*{Community model of the network}
\begin{figure}[!ht]
\begin{center}
\includegraphics[width=0.49\linewidth]{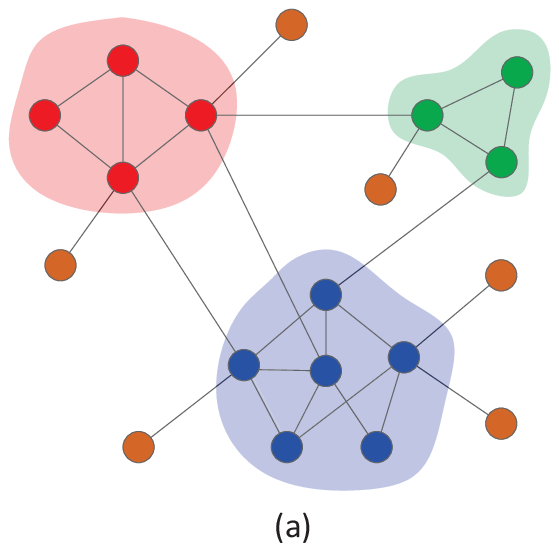}
\includegraphics[width=0.49\linewidth]{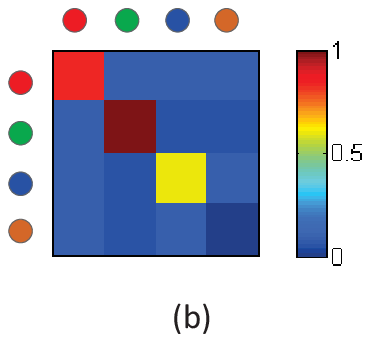}
\end{center}
\caption{An example network to interpret the relationship between community distribution and link density matrix.(a)The community distribution of the network.(b)The link density matrix of the community distribution}
\label{relation2}
\end{figure}

Communities, which are also called modules or clusters, exist widely in real world networks. Intuitively, a community could be a group of nodes with dense connections within a network. Since links tend to crowd in the community, this enlightens us to explore if there exists any underlying correlations between community evolution and link formation and this is also the motivation for us to pursue a new link prediction method based on the distribution of the community structure in a given network. Therefore, We need to analysis the community structure of the network and study the approach to find the communities properly.
\\
Community detection is a fundamental task to exploit the blocks or subgraphs with different properties and functions nested in a network. For social networks, a community could be a group of people with common interest or location. For biology networks, a community could be a group of cells or proteins with common function. Current community detection approaches are "biased" for they are often related to some complex structural features such as sparsity, heavy-tailed degree distribution, and short diameter, etc and also strongly depend on the specific application\cite{25,26,27}. As a result, so far, there is no such a universal measure which is able to determine whether a community obtained by any of these approaches in a given network is true or not fairly. In this paper, we use the measure of link density to quantitively ascertain whether a block is a community or not. If a block $ \mu $ has nodes with number $ |V_{\mu}| $ and edges with number $ |E_{\mu}| $, the link density of the block is defined as follows,
\begin{align}
D_{\mu} = \frac{m_{\mu}}{n_{\mu}}
\label{m1}
\end{align}
Where $m_{\mu}=|E_{\mu}|$ and $n_{\mu}=|V_{\mu}|(|V_{\mu}|-1)/2$. According to Eq. (\ref{m1}), The link density $ D_{\mu} $ denotes the ratio of actual number of inner links to maximal possible number of links in the block $ \mu $. Notice that, when the $ D_{\mu} $ equals 1, the block will reach the highest density and form a complete subgraph which is also called a clique. To quantitively describe the link density between two blocks, we also define the connecting density between every two blocks $ \mu $ and $ \upsilon $ in a given network as follows. Supposing $ |E_{\mu\upsilon}| $ is the number of edges between two blocks while $ |V_{\mu}| $ and $ |V_{\upsilon}| $ are the number of nodes in the block $ \mu $ and block $ \upsilon $ respectively. $ |V_{\mu}| $ multiplying $ |V_{\upsilon}| $ denotes the maximal number of links possiblely existed between the two blocks.
\begin{align}
D_{\mu\upsilon} = \frac{m_{\mu\upsilon}}{n_{\mu\upsilon}}
\label{m2}
\end{align}
Where $m_{\mu\upsilon}=|E_{\mu\upsilon}|$ and $n_{\mu\upsilon}=|V_{\mu}||V_{\upsilon}|$.
Here, we define a community as such a block in a given network which has relatively high inner link density and relatively low connecting density with other blocks. We notice that the structure and characteristic of the communities have naturally exhibited a statistic mechanism of which links are more likely to emerge in the community whereas are less likely to be established between communities. If a network could be partitioned into communities properly, we will be likely to estimate the probability of any node pairs in terms of the distribution of communities.
\\
In Fig. \ref{relation2}(a) , we give an example of network which has been partitioned into three possible communities based on our community definition which are marked with different colors. The conventional community detection methods tend to bind every node in the network to some particular communities enforcedly. But, by doing so, some leaf nodes will be introduced into the community as noise to decreases the inner link density of the community, and this is so-called issue of resolution limit of community detection\cite{28}. And we think this is particularly true in scale-free networks since there are enormous leaf nodes in such kind of network. Unlike the conventional point of view for community detection, we consider that these nodes may not belong to any communities and should be categorized as a group of isolated nodes. Therefore, those leaf nodes which are marked with brown color in Fig. \ref{relation2}(a) are grouped together as a special "community" which has no inner links. Of course, in this "community", links have very little chance to be established among node pairs. If we partition a network into communities in this manner, we can obtain a link density distribution matrix by using Eq. (\ref{m1}) and Eq. (\ref{m2}) to calculate the density within and between the communities. As for Fig. \ref{relation2}(a), we have partitioned the network into four communities including a special "community", thereafter we yield a link density matrix shown in Fig. \ref{relation2}(b).
\\
One network partition can only provide one link density distribution while there usually exist various possible network partitions. If we want to estimate the connecting probabilities for all node pairs in a given network, based on the theory of statistics, we need to obtain independent network partitions as many as possible by doing multiple rounds of network partition. Such procedure is also known as sampling. 
Considering an observed network with a fraction of links removed has adjacency matrix $A^O$, we apply a block patition model $B$ to the observed network. According to the Bayes theorem, the link probability of a node pair $x_{ij}$ can be estimated as
\begin{align}
p(x_{ij}|A^{O})=\dfrac{\sum_{\Omega}\int_{B}p(x_{ij}|A^{O},B)p(A^{O}|B)p(B)dB}{\sum_{\Omega}\int_{B}p(A^{O}|B)p(B)dB},
\label{m3}
\end{align}
where $\Omega$ denotes the space of sampling.
For a node pair $x_{ij}$ in an obtained block $\mu$ by model $B$, we suppose $p(x_{ij}|A^O,b_{\mu})=p_{ij}$. Due to having the high link density in the block $\mu$, it meets that $ {m}_{\mu}\rightarrow{n}_{\mu}$. We have that
\begin{align}
p(A^O|b_{\mu})\approx{p}^{n_{\mu}}_{ij}(1-p_{ij})^{n_{\mu}-m_{\mu}}.
\label{m4}
\end{align}
Let $p(b_{\mu})$ equal a constant. Eq. (\ref{m3}) can be rewritten as
\begin{align}
p(x_{ij}|A^{O})=\dfrac{\sum_{\Omega}\int^{1}_{0}p(x_{ij}|A^{O},b_{\mu})p(A^{O}|b_{\mu})dp}{\sum_{\Omega}\int^{1}_{0}p(A^{O}|b_{\mu})dp},
\label{m5}
\end{align}
Using Eq. (\ref{m4}) and Eq. (\ref{m5}), one can obtain
\begin{align}
p(x_{ij}|A^{O})=\frac{\sum_{\Omega}(n_{\mu}+2)^{-1}\begin{pmatrix}
2n_{\mu}-m_{\mu}+2\\n_{\mu}-l
\end{pmatrix}^{-1}}{\sum_{\Omega}(n_{\mu}+1)^{-1}\begin{pmatrix}
2n_{\mu}-m_{\mu}+1\\n_{\mu}-l
\end{pmatrix}^{-1}}.
\label{m6}
\end{align}
Likewise, for a node pair $x_{ij}$, where node $i$ is in block $\mu$ and node $j$ is in block $\upsilon$, we also suppose $p(x_{ij}|A^O,b_{\mu\upsilon})=p_{ij}$. Due to the low connecting density between the block  $\mu$ and block  $\upsilon$, it meets that $ {m}_{\mu\upsilon}\ll{n}_{\mu\upsilon}$. We have that 
\begin{align}
p(A^O|b_{\mu\upsilon})\approx{p}^{m_{\mu\upsilon}}_{ij}(1-p_{ij})^{n_{\mu\upsilon}}.
\label{m7}
\end{align}
Let $p(b_{\mu\upsilon})$ equal a constant. Eq. (\ref{m3}) can be rewritten as
\begin{align}
p(x_{ij}|A^{O})=\dfrac{\sum_{\Omega}\int^{1}_{0}p(x_{ij}|A^{O},b_{\mu\upsilon})p(A^{O}|b_{\mu\upsilon})dp}{\sum_{\Omega}\int^{1}_{0}p(A^{O}|b_{\mu\upsilon})dp},
\label{m8}
\end{align}
Using Eq. (\ref{m7}) and Eq. (\ref{m8}), one can obtain
\begin{align}
p(x_{ij}|A^{O})=\frac{\sum_{\Omega}(m_{\mu\upsilon}+2)^{-1}\begin{pmatrix}
n_{\mu\upsilon}+m_{\mu\upsilon}+2\\r
\end{pmatrix}^{-1}}{\sum_{\Omega}(m_{\mu\upsilon}+1)^{-1}\begin{pmatrix}
n_{\mu\upsilon}+m_{\mu\upsilon}+1\\r
\end{pmatrix}^{-1}}.
\label{m9}
\end{align}
In the experiments and evaluations section, we mainly use Eq. (\ref{m6}) and Eq. (\ref{m9}) to estimate link probabilities for all node pairs which have no links in the observed network.
\section*{Fast block probabilistic model}
According to the network community model in the previous section, our goal is to partition the network into a set of blocks and ensure each block either is a community or a special “community”, namely a group of isolated nodes having no inner links. It's not trivial to find out all communities in a given network by exhaustive searching because the searching space is usually over large. To fulfill the above task, we proposed a Fast Block probabilistic Model(FBM) by using greedy strategy. Comparing to conventional methods using the rule of Metropolis-Hasting\cite{29}, our algorithm has obtained huge improvment in computing efficiency in accordance with the peformance evaluation results shown in the next section. The FBM Alogorithm is formulated in Table \ref{tab:alg1} and Table \ref{tab:alg2}.
\\
We don't need to provide a mechanism in our algorithm to ensure the relatively low connecting density between blocks because of the fact that almost all real-world networks are sparse network. By implementing our algorithm, we find that it can keep the rare links between obtained every two blocks automatically, i.e., every two blocks has relatively low connecting density. Actually, We have also verified that our algorithm can still work well even through the network is dense. Based on the algorithm, we can quickly partition a given network into communities with high link density and two special communities which are merely grouped by isolated nodes due to the first step in our algorithm of which we initially partition the network into two blocks randomly. Please note this step is essential in our algorithm. Without this step, the network partitions obtained will be strongly correlated. To ensure all network partitions are independent to each other, randomly partition the network into two blocks before implimenting the greedy search of communities is a simple but effective trick in our algorithm which can trigger the procedure of sampling(the independent relationship can be validated by mutual information between partitions). On the other hand, if we remove this step, our algorithm will transform to a pure community detection algorithm. Despite the theoretical and practical interests of community detection, we will not give further investigations on this issue but keep focus on the study of miss link prediction. 

\section*{Experiments and evaluations}
\begin{figure}[!ht]
\begin{center}
\includegraphics[width=0.7\linewidth]{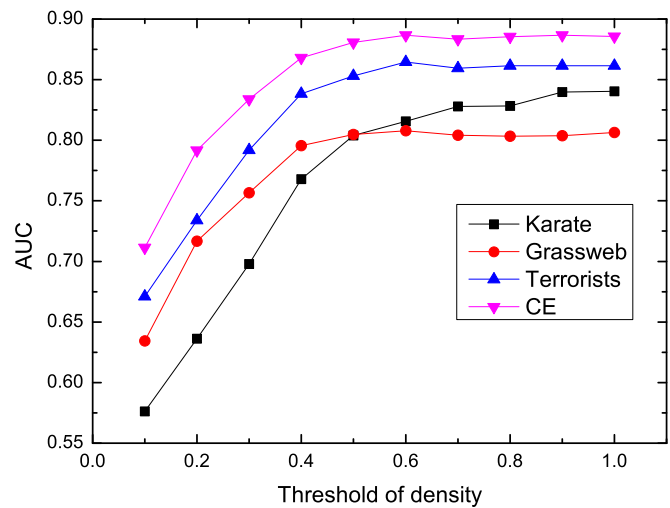}
\end{center}
\caption{Correlations between AUCs and thresholds of link density in the four networks}
\label{threshold}
\end{figure}

\begin{figure}[!ht]
\begin{center}
\includegraphics[width=0.49\linewidth]{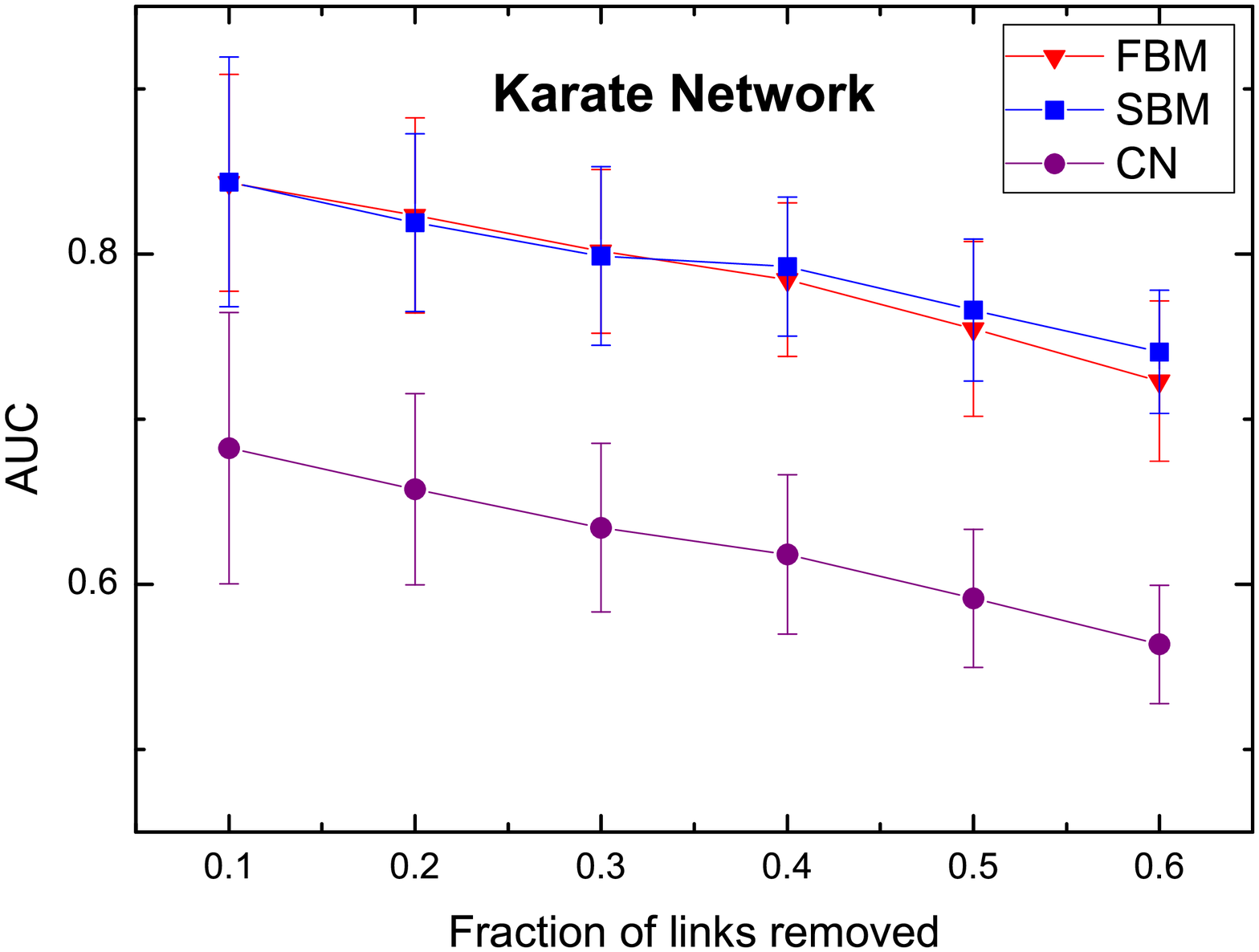}
\includegraphics[width=0.49\linewidth]{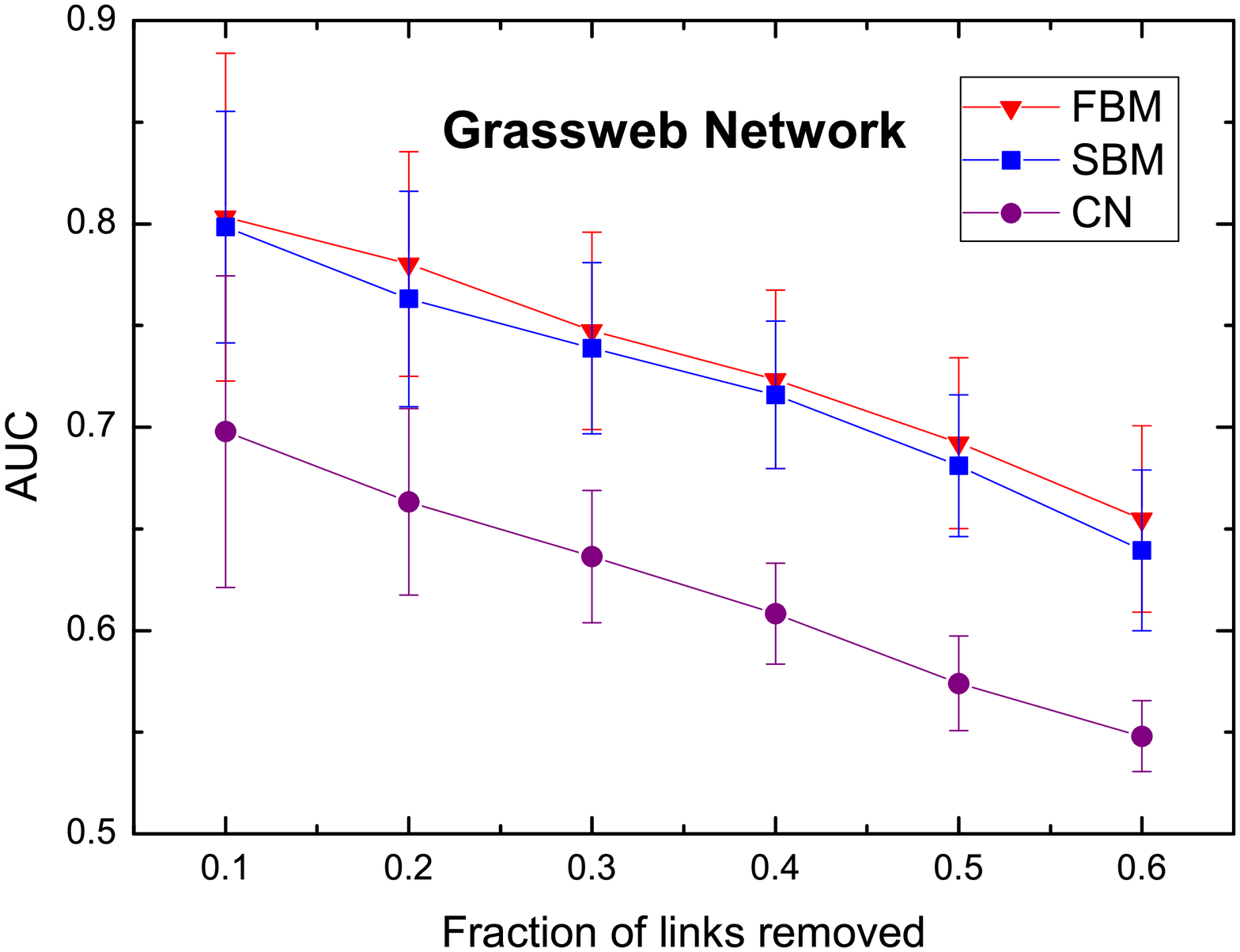}
\includegraphics[width=0.49\linewidth]{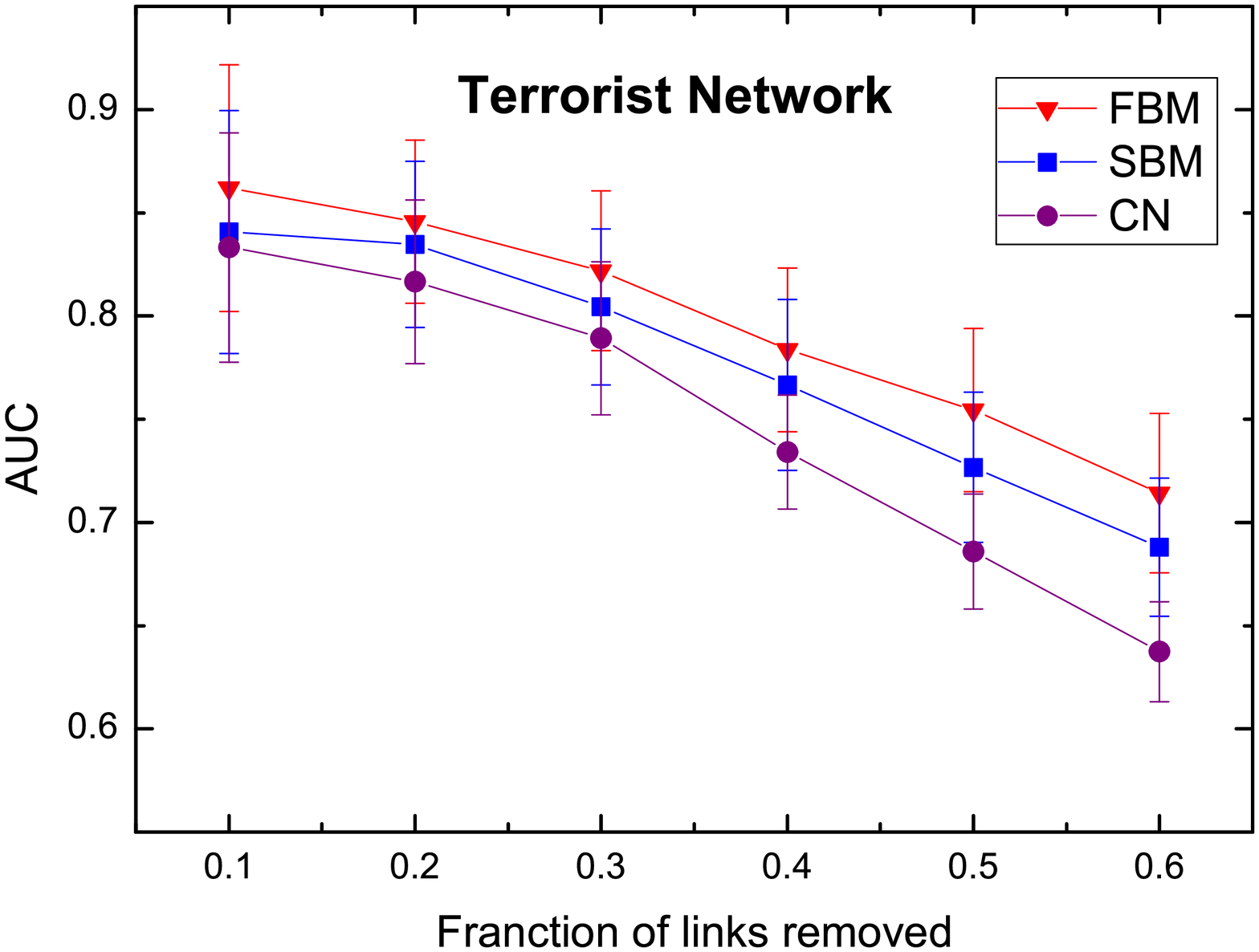}
\includegraphics[width=0.49\linewidth]{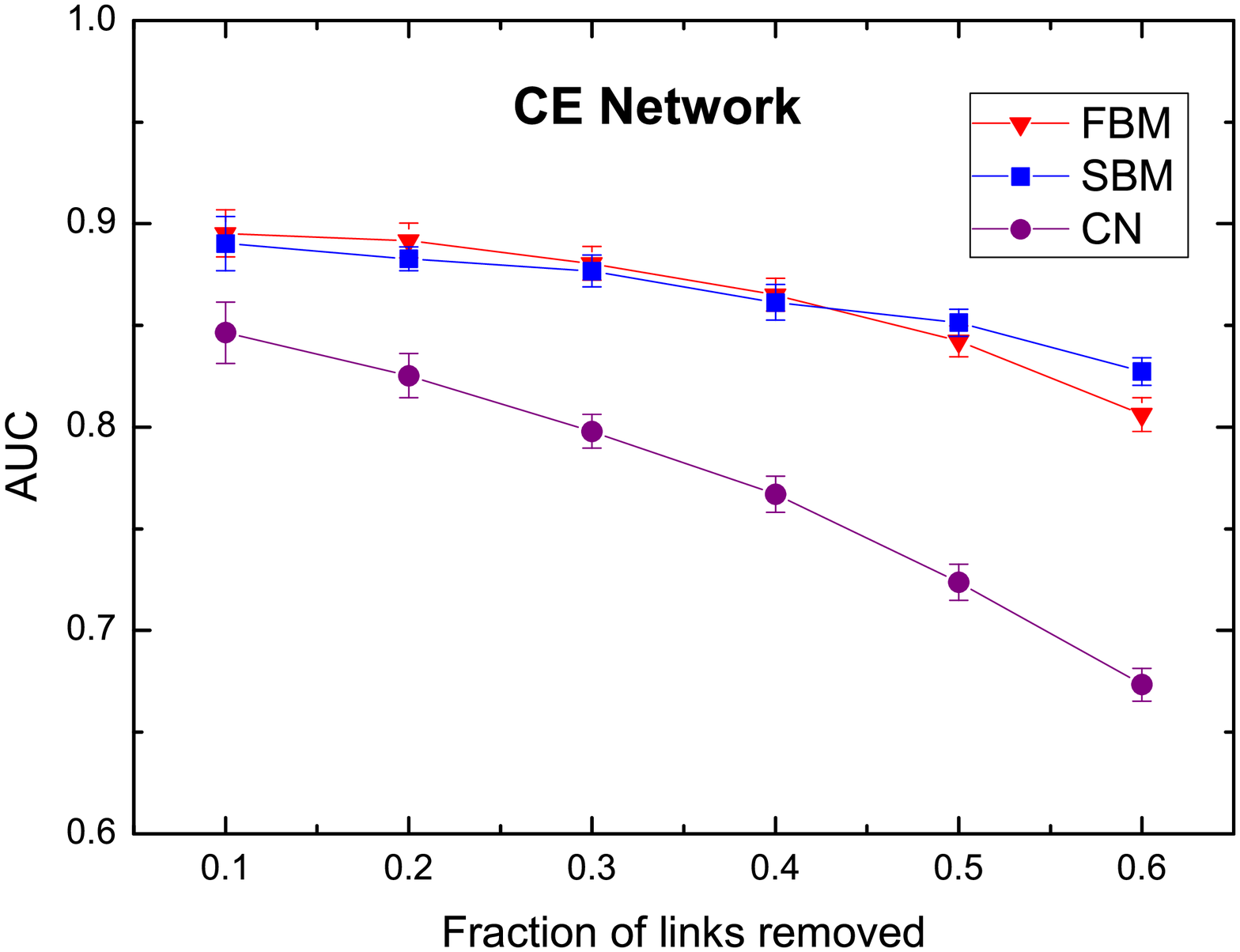}
\end{center}
\caption{Accuracy comparison of link prediction approaches in four networks. Each AUC is avergaed upon 100 implementations and error bar represents the standard deviation.}
\label{auc}
\end{figure}

\begin{figure}[!ht]
\begin{center}
\includegraphics[width=0.49\linewidth]{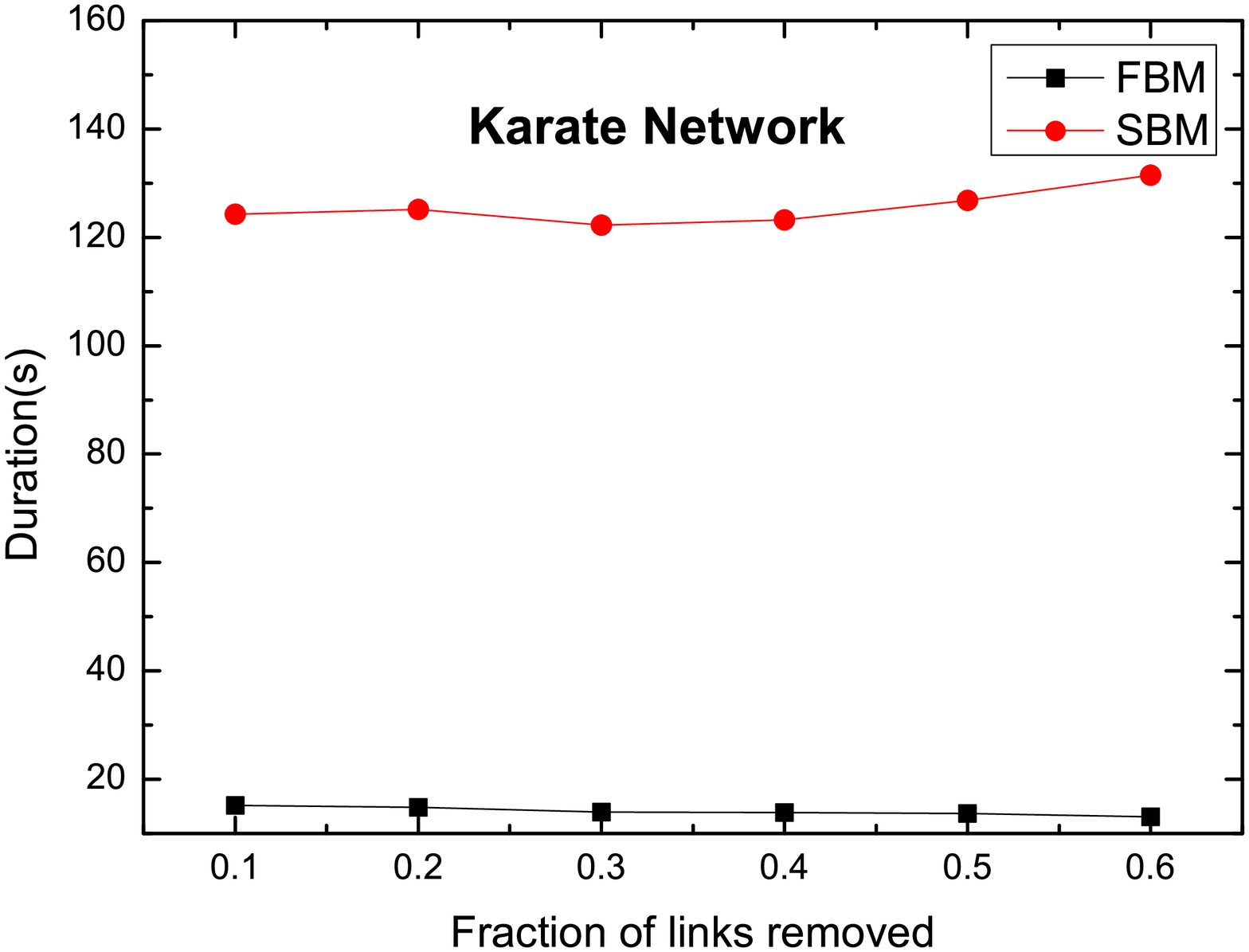}
\includegraphics[width=0.49\linewidth]{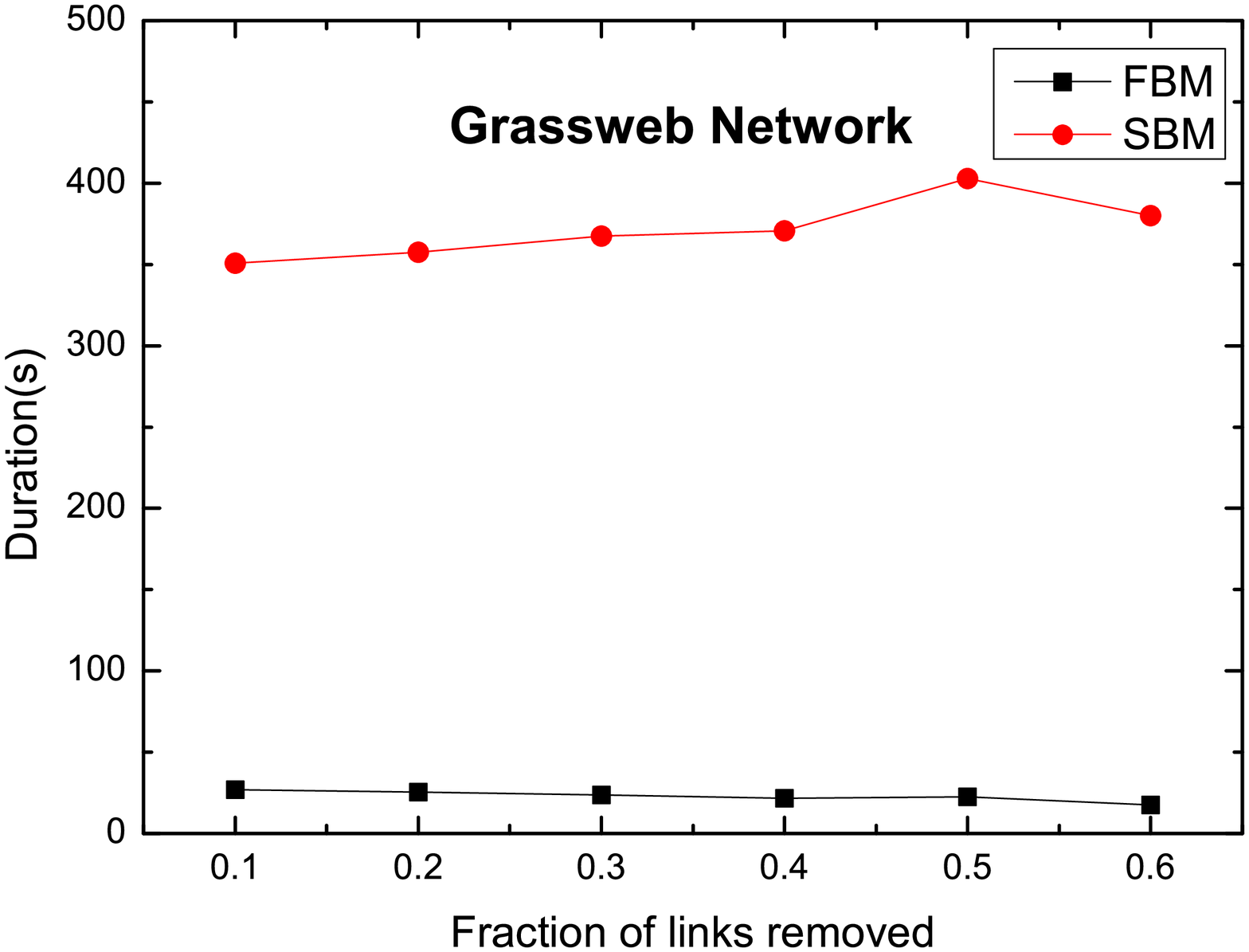}
\includegraphics[width=0.49\linewidth]{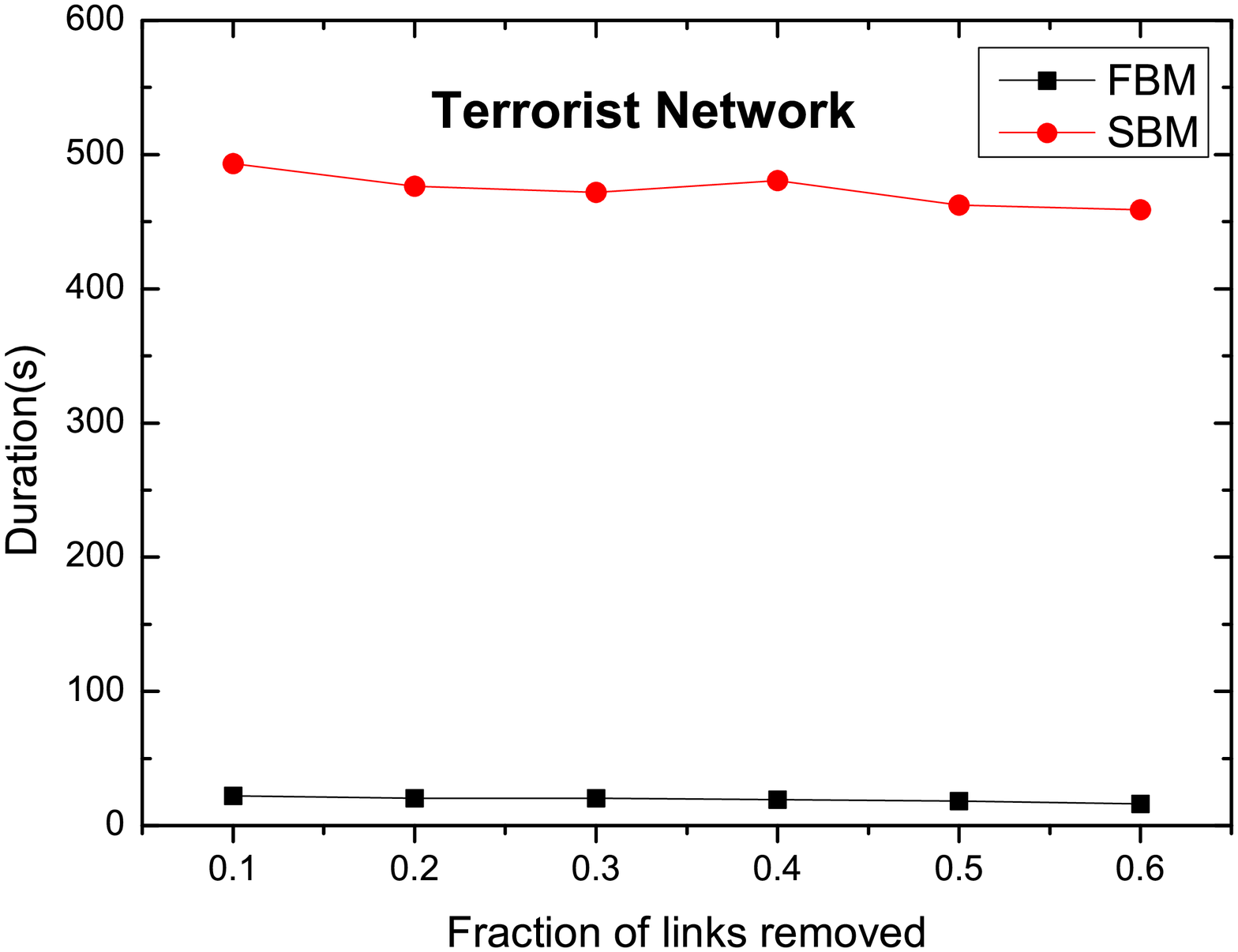}
\includegraphics[width=0.49\linewidth]{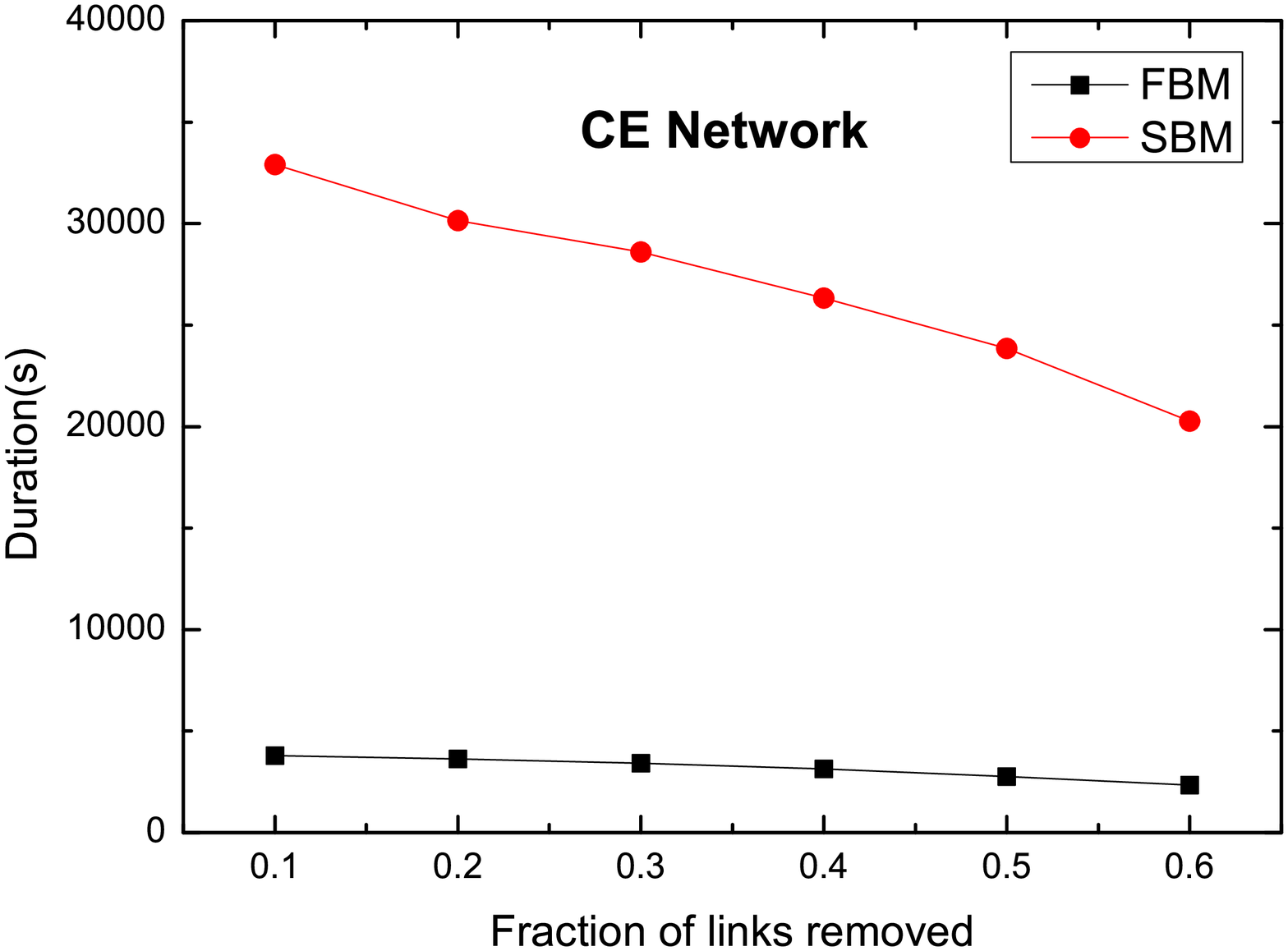}
\end{center}
\caption{Computing efficiency comparison in four networks. Each running time is the lasting period of 100 implementations.}
\label{time}
\end{figure}

In this paper, we consider four real-world networks to implement the tests and evaluations. (1) Social network of friendships between 34 members of a karate club at a US university in the 1970s\cite{30}. (2) Food web of a grassland ecosystem, i.e., a network of predator-prey interactions between species\cite{31}.(3) A network of associations between terrorists\cite{32}. (4)C. elegans (CE): The neural network of the nematode worm C. elegans, in which an edge joins two neurons if they are connected by either a synapse or a gap junction\cite{33}. Here, we only consider the giant component and every network is treated as an indirected network. The topological features statistics of the four networks are summarized in Table \ref{data}.
\\
Before implementing comparisons with other link prediction methods, we still need to determine an uncertain parameter in our alogrithm. As stated in the Table \ref{tab:alg2}, the $threshold$ of the link density must be chosen carefully. We try to pick out the optimized value of the $threshold$ of link density by observing the variation tendency of prediction accuracy along with different link density settings. The measure for prediction accuracy we used here is AUC(area under the receiver operating characteristic curve). AUC can be interpreted as the probability that a randomly chosen missing link is given a higher score than a randomly chosen non-existent link. Fig. \ref{threshold} shows the accuracy variation curves ploted for the four networks as the fraction of missing links is set to ten percent\footnote{we obtained similar results as other different fractions of missing links are set to.}. We found the accuracy of link prediction tends to coverage after the link density is larger than 0.5 and reaches the best when the $threshold$ of link density is set to 1 while the block corresponds to a clique.
\\
To estimate the likelihood of missing links, researchers have developed various probabilistic prediction models in recent years. A typical model, Hierarchical Random Graph(HRG), proposed by Aaron Clauset et al, was applied to predict miss links in some networks with obvious hierarchical structure\cite{8}. Based on the similar theory of statistics using by HRG  model but from another angle of view, Roger Guimera et al proposed a Stochastic Block Model(SBM) which can predict both missing links and spurious links and is able to give a much better accuracy of prediction in various kind of networks compared to some other popular methods including the HRG\cite{10} approach. To our best knowledge,the SBM algorithm is the state-of-the-art approach with the best accuracy of prediction in indirected network without weight information.
\\
We mainly made performance comparisons both on missing link prediction accuracy and computing efficiency between our algorithm and the SBM approach. And the accuracies of the common neighbor method are presented here as a baseline. Our algorithm and the SBM approach have a common characteristic which are both required to sample network partitions. To ensure the comparison is fair to the both approaches, we apply the same sampling standard to them which is set to 50 times. The hardware we use to test is a desktop with processor of Intel(R) Core(TM) i7 CPU 930 @ 2.8Ghz(eight cores) and 8 GigaBytes memory.
\\
The prediction accuracies, measured by AUC for the four networks, are plotted in Fig. \ref{auc} and the corresponding comparisons of running time(the unit is second) implimented in the four networks are shown in Fig. \ref{time}. To ensure the results are trusted, each value of accuracy is obtained by averaging over 100 implementations with independently random network divisions of training set and probe set while the error bars denote the standard deviation. Accordingly, each value of the running time is the lasting period over 100 implementations.
\\
According to the AUC comparisons shown in Fig. \ref{auc}, the FBM approach performs better than the SBM approach in the networks of Grassweb and Terrorist, and has very close accuracy result to that of the SBM approach in the other two networks. Meanwhile, According to the computing efficiency comparisons shown in Fig. \ref{time}, the running time used by the FBM approach are far less than that used by the SBM approach. During the experiments, we found that the running time consumed by the SBM approach increases rapidly along with the size growth of the network while that of the FBM approach increases mildly, and this indicates that the SBM algothim has much high time complexity than the FBM algorithm and is also the main reason that why we have chosen the networks with relatively small size to make the comparisons. The experiment results prove that the FBM approach is able to give very good accuracy for missing link prediction on real world networks with superior computing efficiency.

\section*{Mechanism analysis of the link formation}
\begin{figure}[!ht]
\begin{center}
\includegraphics[width=0.31\linewidth]{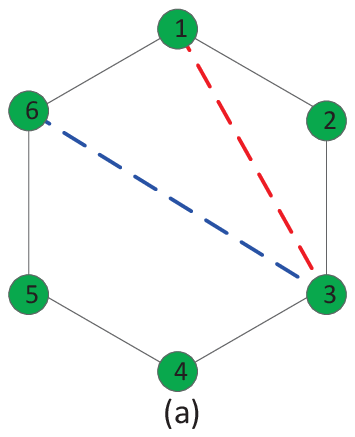}
\includegraphics[width=0.31\linewidth]{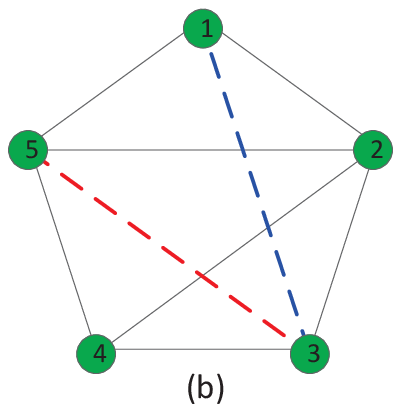}
\includegraphics[width=0.31\linewidth]{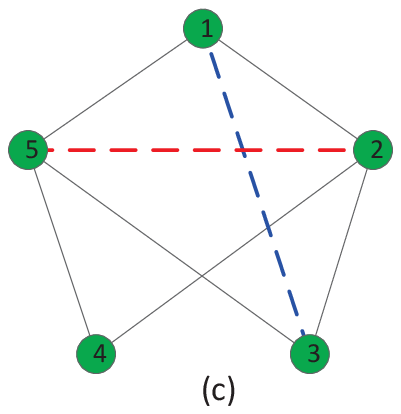}
\end{center}
\caption{Three artificial networks to interpret the link formation mechanism in the communities}
\label{mech}
\end{figure}

To find out the reason that why the FBM approach is of the capacity to give very good accuracy for missing link prediction in real world networks, we try to analyze what kind of links have higher link probability by investigating the probabilistic distributions of node pairs in the four networks we have tested. We observed there are mainly three important principles which may build the theoretical basis to explain what kinds of links are likely to emerge in the networks. To demonstrate the three principles easily, we give three example networks shown in Fig. \ref{mech}, each of which is likely to uncover one kind of links favored by the FBM model. 
\\
Fig. \ref{mech}(a) shows a ring structure with six nodes labeled by numbers and two possible links between node pair (1,3) and node pair (3,6) which are denoted by red dash line and blue dash line. We evaluate the likelihoods of the two possible connections by applying the FBM approach to the network. The results show that the connection probability of node pair (3,6) is only 50 percent of that of node pair (1,3) which means that the node pair (1,3) is more likely to connect together compared to node pair (3,6). We notice that, if a link added between node pair (1,3), a clique (1,2,3) would be established. This indicates that link tends to in priority establish a clique in a network. Fig. \ref{mech}(b) shows a five-node network and two possible links between node pair (1,3) and node pair (3,5) which are also denoted by blue dash line and red dash line respectively. After calculating the link probabilities of the two node pairs, we found that the connection probability of node pair (1,3) is 75 percent of that of node pair (3,5). The difference of the two connection probabilities reminds us that an addition of link (3,5) will form a larger clique (2,3,4,5) than the clique (1,2,3) if link (1,3) added. This case demonstrates that link tends to create larger clique first in a network when there are many options available to choose. Fig. \ref{mech}(c) shows another interesting phenomenon of link formation. After calculation, the link probabilities of node pair (3,5) is 1.5 times higher than that of node pair (1,3). We found that link (3,5) could create three cliques including (1,2,5),(2,3,5) and (2,4,5) while link (1,3) would only be able to create two cliques, i.e., (1,2,3) and (1,3,5). This result implies that if adding a link is able to create more cliques, the link will have higher likelihood to be established. In terms of the three typical cases, we summarized three principles to explain how link creates.
\\
(i)Link is very likely to be established to form a clique in a given network.
\\
(ii)Link prefers to create larger clique to smaller clique in a given network.
\\
(iii)Link tends to form cliques as many as possible in a given network.
\\
As stated in section 2, links tend to crowd in the communities. We believe the three principles would be reasonable to explian the link formation mechanism in the communities. To prove this hypothesis, we made an addtional expriment on the four networks which have been used to test in the prior section. We firstly set the link density $threshold$ to 0.8 and apply the FBM approach to partition each network into communities respectively. Then we remove 10 percent of links existed in the communities by following the three princples as probe set and keep the rest of the network as training set. We still use the AUC to evaluate the accuracy of prediction and obtain the average results upon 100 time implementations shown in Table \ref{tab:auc}. Comparing to the results shown in the prior section, we found that the new predicition results are even better which prove that the FBM has definitely applied the three principles to predict the missing links within the communities. In other word, it has also proved that the three principles can well capture the essence of link formation and reveal the rule of community growing and evoluation in real world networks.
\\
The mechanism of common neighbor approach is usually explained by the social balance theory\cite{34,35}, but it is actually a special case which has applied the principle (iii), since if a given node pair has many common neighbors, it also means that it would form many triangle cliques after a link is added to the node pair. Therefore, this approach can still perform well in some cases. But due to only partial essense of link formation captured by this method, it could not give good accuracy results in our experiments.

\section*{Discussion}
The three principles have suggested that the clique stucture plays a significant role to drive link formation in the communites. In this section, we want to give more empirical evidences provided by other researcher to discuss why clique can fulfil this task. Compared to other structures, such as asterial structure and ring structure, the clique has some unique features. Firstly, the clique is the most complete structure which has the most dense links and hence ensures it is the most stable and robust component of the network. Secondly, the clique has shortest distance between every two nodes which ensure it has very high efficiency for communication. It has been revealed that, in Internet network, some hub nodes like backbone routers tend to connect together which is so-called rich-club phenomenon\cite{36}. The rich-club will improve the efficiency of traffic routing and provide the capacity to resist some node attacks and prevent the network from breaking down easily. As rich-club is a typical community in terms of our definition, the clique might have the underlying impact to promote the rich-club's formation and growth in the network. It’s also found in biology network that motifs often have the structure of the clique, such as the feed-forward loop, which is known as directed triangle motif, emerges in both transcription-regulatory and neural networks\cite{37}. Meanwhile, the research finding has revealed that these motifs tend to cluster together which exhibit as a general property in all real networks, so, new links tend to emerge during this clustering procedure. The above known research achievements in different domains have provided more solid empirical evidences to support our theory.
\\
Previous studies have also verified that the local clustering property such as node’s clustering coefficient can be utilized to improve the accuracy of link prediction, yet they did not give any solid reason about their methods[19]. Clique is a typical stucture which is of the property of local clustering in the network. Our work has revealed that such structure is the important cradle to promote link formation in community and network evolution. With an intuitive insight, we also believe that our model has the potential to provide some new evidence to explain why so many networks in the real world exhibit the topology of hierarchy communities if further network evolution study can be done under the link formation mechanism we've found. 
\section*{Conclusions}
In this paper, we proposed a Fast probability Block Model(FBM) to predict missing links in complex networks. In terms of the experiment results in four real-world networks, the FBM model has exhibited slight better accuracy performance and overwhelmed better computing efficiency than the state-of-the-art model SBM. We believe the FBM approach has the potential to give fairly good accuracy of link prediction in much larger and more complex networks such as massive biology networks, rapid growing social networks, and World Wide Web networks, etc. So, compared to the SBM approach, The FBM algorithm is more applicable. On the other hand, from the theoretical aspect, we revealed that network’s clique structure plays an important role to drive link formation and community evolution. And the underlying mechanism of link formation in communities have been well interpreted by three principles summarized in this paper which can provide researchers a new framework for link prediction study. Meanwhile, our model is very likely to give enhanced prediction accuracy in specific applications when domain knowledge is introduced such as node properties and edge features. furthermore, FBM model is of a good outlook to give an insight on exploring new methodology of community detection in complex networks although this is not investigated deeply in this paper.

\section*{Acknowledgments}
 This work is partially supported by the national natural science foundation of China under grant Nos. 60903073, 61103109 and HuaWei(No. YBCB2011057) university-enterprise cooperation project and the central university research foundation under grant No. ZYGX2012J085.

\bibliography{plos2009}

\section*{Figure Legends}
\section*{Tables}
\begin{table}[!ht]
\caption{
\bf{The algorithm of Fast Block probabilistic Model}}
\begin{tabular}{|l|}
\hline
Input: a network $G(V,E)$;\\
Output: communities $(C_1,C_2,...,C_m)$ of the network $G$;\\
\hline
1.	Let $G_1=(V_1,E_1)$ and $G_2=(V_2,E_2)$ where $V=V_1\cup{V_2}$, $\varPhi= V_1\cap{V_2}$ and $\varPhi= E_1\cap{E_2}$;\\
/*the blocks $G_1$ and $G_2$ are partitioned randomly by the network $G(V,E)$*/\\
2.	For each $G_i$ do\\
3.	~~  $j$=1;\\
4.	~~  While $E_i<>\phi$ do /* loop stops when no edges exist in $G_i$*/\\
5.	~~~~~~      $CommunityFind(G_i,C_{j})$; /*procedure to find a community $C_{j}$ from $G_i$ with high density*/\\
6. ~~~~~~	      Output($C_{j}$);\\
7. ~~~~~~	      $G_i$=Remove($C_{j}$, $G_i$); /*remove the community $C_{j}$ from $G_i$ including nodes and \\~~~~~~~~~~~~edges which belong to $C_{j}$ and interconnections between $C_{j}$ and $G_i$ */\\
8.	~~~~~~  $j$=$j$+1;\\
9.	~~	End while\\
10.	~  Output($G_i$); /*the remaining graph $G_i$ will be a group of isolated nodes*/\\
11.	End for\\
\hline
\end{tabular}
\label{tab:alg1}
\end{table}

\begin{table}[!ht]
\caption{
\bf{The procedure of CommunityFind called by the algorithm of Fast Block probabilistic Model}}
\begin{tabular}{|l|}
\hline
Input: a network $G_i$;\\
Output: a community $C_j$;\\
\hline
1.	$A$=Density($G_i$); /*calculate the density of the network $G_i$*/\\
2.	while  $A<threshold$ do  /*$threshold$ is an accepted maximum value of link density */\\
3.	~~     Sort($V_i$); /*sort all nodes by node degree in descending order*/\\
4.	~~     $G_i$=Remove($v$,$G_i$); /*remove the node $v$ in the top of the list with the least degree and edges \\
~~~~~~	                       attached to the $v$ from $G_i$, and derive a new network $G_i$*/\\
5.	~~     $A$=Density($G_i$); /*recalculate the density of network $G_i$*/\\
6.	End while\\
7.  $C_j$=$G_i$;\\
8.	Return($C_j$);  /*derive a community $C_j$ with $threshold$ density */\\
\hline
\end{tabular}
\label{tab:alg2}
 \end{table} 
 
\begin{table}[!ht]
\caption{
\bf{Topological features statistics of the four networks. $|V|$ and $|E|$ are the number of nodes and links. $C$ and $D$ are clustering coefficient and density of network, respectively. $M$ is the modularity of network. $\langle{k}\rangle$ and $\langle{d}\rangle$ are the average degree and the average shortest distance.}}
\begin{tabular}{|c|c|c|c|c|c|c|c|}
\hline
           &       $|V|$  &        $|E|$  &       $C$  &        $D$ &        $M$ &      $\langle{k}\rangle$ & $\langle{d}\rangle$ \\
\hline
    Karate   &  34	&  78	&  0.588	&  0.139	&  0.416	&  4.588	&  2.408	\\
   Foodweb   &  75	&  113	&  0.497	&  0.041	&  0.635	&  3.013	&  3.875	\\
 Terrorists   &  62	&  152	&  0.58  	&  0.08	    &  0.529	&  4.903	&  2.508    \\
        CE   &  297	&  2148	&  0.308	&  0.049	&  0.397	&  14.465	&  2.946    \\
\hline
\end{tabular}
\label{data}
 \end{table}

\begin{table}[!ht]
\caption{
\bf{The accuracy results of missing link predcition in communitis of the four networks. The threshold of link density to partition communites is set to 0.8 for each network. And the fraction of links removed from the communities is 10 percent.}}
\begin{tabular}{|c|c|c|c|c|}
\hline
  Network & Karate   &  Foodweb	&  Terrorists	&  CE	\\
\hline
   AUC   &  0.9427$\pm$0.0583	&  0.9013$\pm$0.0471	&  0.9585$\pm$0.0387	&  0.9535$\pm$0.0076	\\
\hline
\end{tabular}
\label{tab:auc}
 \end{table}
\end{document}